\documentclass [11pt]{article}
\pdfoutput=1
\usepackage{amsfonts}
\usepackage{graphicx}
\usepackage{amsmath,amssymb, dsfont}
\usepackage{latexsym}
\usepackage{mathrsfs}
\usepackage{bbold}
\usepackage{color}
\usepackage{slashed}
\usepackage{cancel}
\usepackage{soul} %
\usepackage{setspace} %
\usepackage{comment}
\usepackage{booktabs}
\usepackage{multirow}
\usepackage{cite}
\usepackage[small]{caption}

\definecolor{red}{rgb}{1,0,0}

\makeatletter
\def\section{\@startsection {section}{1}{\z@}{-3.5ex plus -1ex minus
 -.2ex}{2.3ex plus .2ex}{\large\bf}}
\def\subsection{\@startsection{subsection}{2}{\z@}{-3.25ex plus -1ex
minus -.2ex}{1.5ex plus .2ex}{\normalsize\bf}}
\makeatother
\makeatletter

\@addtoreset{equation}{section}

\makeatother

\newcommand{\bea}{\begin{equation} \begin{aligned}} \newcommand{\eea}{\end{aligned} \end{equation}}
\def\be{\begin{equation}} \def\ee{\end{equation}}

\usepackage[colorlinks,linkcolor=black,citecolor=blue,urlcolor=blue,linktocpage,pagebackref]{hyperref}

\allowdisplaybreaks

\begin{document}

\thispagestyle{empty}

\begin{center}

	\vspace*{-.6cm}

	\begin{center}

		\vspace*{1.1cm}

		{\centering \Large\textbf{Five-dimensional CFTs from the $\epsilon$-expansion}}

	\end{center}

	\vspace{0.8cm}
	{\bf Fabiana De Cesare$^{a}$, Lorenzo Di Pietro$^{b,c}$ and Marco Serone$^{a,c}$}

	\vspace{1.cm}
	
	${}^a\!\!$
	{\em SISSA, Via Bonomea 265, I-34136 Trieste, Italy}

	\vspace{.3cm}

	${}^b\!\!$
	{\em  Dipartimento di Fisica, Universit\`a di Trieste, \\ Strada Costiera 11, I-34151 Trieste, Italy}
		
	\vspace{.3cm}

	${}^c\!\!$
	{\em INFN, Sezione di Trieste, Via Valerio 2, I-34127 Trieste, Italy}

	\vspace{.3cm}

\end{center}

\vspace{1cm}

\centerline{\bf Abstract}
\vspace{2 mm}
\begin{quote}

We look for UV fixed points of non-abelian $SU(n_c)$ gauge theories in $4+2\epsilon$ dimensions with $n_f$ Dirac fermions in the fundamental representation, 
using the available five-loop $\overline{{\rm MS}}$ $\beta$-function and employing Pad\'e-Borel resummation techniques and Pad\'e approximants to the series expansion in $\epsilon$.
We find evidence for a $5d$ UV-fixed point for $SU(2)$ theories with $n_f\leq 4$. % and pure $SU(n_c)$ theories for $n_c\leq 4$. 
We also compute the anomalous dimensions $\gamma$ and $\gamma_g$ of respectively the fermion mass bilinear and the gauge kinetic term operator at the UV fixed point.
%Similar results are obtained by means of ordinary Pad\'e approximants.

\end{quote}

\newpage

\tableofcontents

\section{Introduction}
Conformal Field Theories (CFTs) play a major role in theoretical physics.  They are the starting and ending points of renormalization group (RG) flows in quantum field theories,
they describe second-order phase transitions in critical phenomena and they possibly allow us to have a non-perturbative definition of quantum gravity theories by means of the AdS/CFT correspondence. In absence of extra symmetries, most notably  supersymmetry, finding CFTs become increasingly difficult as the dimension of space-time $d$ increases. The only analytical evidence of the existence of $4d$ non-supersymmetric CFTs comes from the Veneziano limit (large number of colors and flavors) of non-abelian gauge theories, where we can tune the one-loop coefficient of the $\beta$-function to be parametrically small and negative, while having a positive two-loop coefficient (Caswell-Banks-Zaks \cite{Caswell:1974gg,Banks:1981nn} fixed points). To what extent the IR fixed point persists at finite $N$ is a non-trivial question which still has to be settled.

In $d>4$, gauge theories are perturbatively non-renormalizable and should be considered effective field theories which become free theories in the IR.
Therefore in $d>4$ the natural question is whether there exists a UV fixed point, i.e. an interacting CFT that when deformed by a relevant operator admits an effective description as a non-abelian gauge theory. The existence of such a fixed point for Yang-Mills theories in $d>4$ would be analogous to well-known lower-dimensional examples of perturbatively non-renormalizable theories with a non-trivial continuum limit, such as the non-linear $\sigma$-model. This theory is renormalizable and asymptotically free in two dimensions and has a non-trivial UV fixed point which can be studied with the $\epsilon$-expansion in $2<d<4$. Since this model belongs to the same universality class of $\lambda\varphi^4$ scalar theory, the corresponding critical behaviour can be verified by matching the results obtained applying the $\epsilon$-expansion to both theories (in $d=2+\epsilon$ and $d=4-\epsilon$). A similar story holds for the Gross-Neveu theory in $2<d<4$.

A parametrically weakly coupled UV fixed point in non-abelian gauge theories can in fact be established in $4+2\epsilon$ dimensions with $\epsilon \ll 1$ \cite{Peskin}.
%\footnote{The idea is identical to that used to for the Caswell-Banks-Zaks trick, replacing one and two loop terms with tree-level and one-loop in the $\beta$-function, with reversed signs.} 
 It is of course crucial to understand if this UV fixed point persists up to $d=5$. Evidence that this is the case has been provided by 
\cite{Morris_2005}, where fixed points of this kind have been studied up to ${\cal O}(\epsilon^4)$. The analysis in \cite{Morris_2005} used the then-available four loop $\beta$-function for non-abelian gauge theories and was based on the optimal truncation of the series in $\epsilon$.  Note that no example of non-supersymmetric interacting unitary CFTs is currently known in $d=5$. A conjectural example was the UV fixed point of the $O(N)$ model, whose $1/N$ expansion was shown in \cite{Fei:2014yja} to be compatible with unitarity for $N$ larger than a certain critical value $N_c$. However it was later realized that this fixed point is rendered non-unitary even for  $N>N_c$ by certain instantonic contributions to the imaginary part of observables, that are exponentially small at large $N$ \cite{Giombi:2019upv}. 

The aim of this paper is to extend the analysis of \cite{Morris_2005} by adding one more term in the analysis, thanks to the by-now known five loop $\beta$-function \cite{Baikov:2016tgj,Herzog:2017ohr,Luthe:2017ttg,Chetyrkin:2017bjc}, and to
use Borel resummation techniques, which allow us to have better control on the (plausibly) asymptotic series in $\epsilon$.
We consider $SU(n_c)$ gauge theories  with $n_f$ Dirac fermions in the fundamental representation of the gauge group.
We find evidence for the existence of UV fixed points when both $n_c$ and $n_f$ are small enough. The evidence gets weaker and weaker as either $n_c$ or $n_f$ increases.
%We find good evidence for the existence of UV fixed points when both $n_c$ and $n_f$ are small enough. The evidence gets weaker and weaker as
%either $n_c$ or $n_f$ increases, so the most reliable result applies for pure SU(2) non-abelian gauge theories. \FDC{Questo non è completamente vero, nel senso che è vero che la banda è più piccola per $n_f=0$, ma sta un po' più sopra per $n_f=1,2,3$. Siccome era così anche prima (cft vecchia versione), questa frase la lasciamo? }
Very similar results are obtained using ordinary Pad\'e approximants, without using Borel resummation techniques.

The paper is structured as follows. In section \ref{sec:stage} we briefly set the stage of our computation, which follows the same logic used in \cite{DiPietro:2020jne} to study the conformal window in $4d$ QCD. We present our results in section \ref{sec:results}, which include also the computation of the anomalous dimension $\gamma_g$ of the gauge kinetic term operator ${\rm tr} [F_{\mu \nu} F^{\mu \nu}]$ and of the fermion mass operator  $\gamma$ (when present) $\bar \psi \psi$.
The existence of a $5d$ UV fixed point in non-abelian gauge theories is debated in the literature.
We briefly review in section \ref{sec:overview} previous studies  by means of different techniques aimed at looking for $5d$ non-supersymmetric CFTs.
We summarize our results and conclude in section \ref{sec:conclu}.

\section{Preliminaries}
\label{sec:stage} 

The existence of a Wilson-Fisher fixed point for Yang-Mills theories in $d=4+2\epsilon$ with $\epsilon\ll 1$ is easily established. 
By denoting the loopwise expansion parameter 
\be
a \equiv \frac{g^2}{16\pi^2}
\ee
where $g$ is the usual gauge coupling constant, we have in $d=4+2\epsilon$ dimensions
\begin{equation}
\beta(a)=\frac{1}{2}\frac{da}{d\log \mu}=\epsilon a+\beta^\mathrm{4d}(a) \equiv  \epsilon a -\sum_{n=0}^\infty \beta_n \,a^{n+2}\,,
\label{eqn:46}
\end{equation}
where $\beta^\mathrm{4d}$ denotes the ordinary $4d$ $\beta$-function.
For $n_f<\frac{11}{2}n_c$ the leading contribution to $\beta^\mathrm{4d}(a)$ is famously negative and hence the existence of a parametrically weakly coupled 
UV stable fixed point can be established for $\epsilon \ll 1$, as pointed out long ago by Peskin \cite{Peskin}. The question of whether or not this result can be extended to a physical number of dimensions, in particular $d=5$ ($\epsilon=0.5$), requires an analysis of higher order terms in $\beta^\mathrm{4d}$. This beta function, as well as the fermion mass anomalous dimension $\gamma$, for generic gauge groups with $n_f$ fundamental fermions is known up to five-loop orders \cite{Baikov:2016tgj,Herzog:2017ohr,Luthe:2017ttg,Chetyrkin:2017bjc,Baikov:2014qja,Luthe:2016xec,Baikov:2017ujl} in $\overline{{\rm MS}}$.\footnote{General expressions for $\beta^{{\rm 4d}}(a)$ and $\gamma(a)$ can be found in, e.g.,  \cite{Herzog:2017ohr} and  \cite{Baikov:2017ujl}, respectively.} As is well-known, in quantum field theory the loopwise expansion of physical observables is generally divergent asymptotic.
In non-abelian gauge theories the asymptotic expansion is also non-Borel resummable because of  instantons and renormalon singularities.
On the other hand, the nature of the coupling expansion of non-physical quantities, like the $\beta$-function, depends on the renormalization scheme.
It is still unclear whether $\beta$ in $\overline{{\rm MS}}$ is convergent or divergent asymptotic and, in the latter case, if the associated series is Borel resummable or not. 

Theoretical and numerical arguments supporting a non-Borel resummable nature of the series for $\beta$ and $\gamma$ have been given in \cite{DiPietro:2020jne}, where it has been pointed out that the
perturbative $\beta$-function could miss non-perturbative contributions of the kind\footnote{{\textbf{Note added (August 2024)}: With the notation for the $\beta$-function in eq.\eqref{eqn:46}, the dynamically generated scale $\Lambda_\text{QCD}$ gets a factor 1/2 in the exponent. This factor was missed in previous versions, leading to an incorrect value for the non-perturbative error. This correction compels us to be less conservative in the choice of the arbitrary parameter $c_\text{np}$ defined in eq.\eqref{eqn:51}. Consequently,  we take here $c_\text{np}$ to be 1 instead of 10, as was  done in previous versions.We note that the evidence for some of our results is reduced. } }
\begin{equation}
\delta\beta^{{\rm 4d}}\sim  \hat{h} \left(\frac{\Lambda_{\text{QCD}}}{\mu}\right)^k = \hat{h} \,e^{-\frac{k}{{2\beta_0 a}}}~\,,
\label{eq:betaNP}
\end{equation}
with $k\geq 2$. If the series expansion of $\beta^{{\rm 4d}}(a)$ is divergent, so is also the series of the function $\epsilon(a^*)$ obtained by solving $\beta(a^*) = 0$, and its inverse
$a^*(\epsilon)$: 
\begin{equation}
a^*(\epsilon)=\epsilon\sum_{n=0}^\infty b_n \epsilon^n, \quad \beta(a^*)=0.
\label{eqn:48}
\end{equation} 
We will assume in the following the most conservative and worst-case scenario, namely that the coupling expansion of $\beta(a)$ is divergent asymptotic and non-Borel resummable. As explained in \cite{DiPietro:2020jne}, depending on the location of the leading singularity in the Borel plane, the numerical reconstruction of a formally non-Borel resummable function might lead to a better accuracy in the ending result compared to perturbation theory or optimal truncation.
We use Pad\'e approximants to estimate the Borel transform of $a^*(\epsilon)$ as well as of $\gamma^*=\gamma(a^*)$ and $\gamma_g^* = \gamma_g(a^*)$.
This method is commonly dubbed as Pad\'e-Borel resummation.\footnote{Our knowledge of the analyticities properties of the Borel function associated to these observables is unfortunately too
limited to implement more efficient resummation techniques based on conformal mappings.}
In order to avoid confusion with the use of Pad\'e approximants that directly estimate a quantity rather than its Borel transform,
we will denote by Pad\'e-Borel approximants those entering in the reconstruction of the Borel transform, keeping the name of Pad\'e approximants for the direct estimate of the quantity.
The numerical implementation of our procedure essentially follows that used in \cite{DiPietro:2020jne} to study the QCD conformal window. 
See appendix \ref{appendix} for details.

The existence of the fixed point $a^*$ will be considered reliable only if the error band does not reach negative values, which would correspond to a possibly unphysical fixed point.  \\
Since we do not know whether the series in $\epsilon$ are convergent or divergent asymptotic, we are not guaranteed to do better with Pad\'e-Borel rather than simple Pad\'e approximants.
Theoretically, namely by knowing a parametrically large number of perturbative coefficients, we would expect to better reconstruct a quantity using simple Pad\'e approximants if that is analytic
at $\epsilon=0$, or Pad\'e-Borel approximants if that is non-analytic at $\epsilon=0$. In practice, having only a few perturbative coefficients at our disposal, such considerations cannot be tested.
For this reason, we have also considered ordinary Pad\'e approximants for $a^*$, $\gamma^*$ and $\gamma_g^*$, defining an error band to each approximant. This error is identical 
to that used in the Pad\'e-Borel method but does not contain the contribution \eqref{eqn:a8} in appendix \ref{appendix}. 
The results obtained by taking ordinary Pad\'e approximants and by Pad\'e-Borel resummations are nicely in remarkable agreement. The central values are essentially identical and often the error bands are 
very close. Only in a few cases Pad\'e-Borel resummations give more accurate results. For this reason we will report in section \ref{sec:results} only the results obtained using the Pad\'e-Borel method.

\begin{figure}[t!]
 \centering
\includegraphics[height=0.294\textwidth]{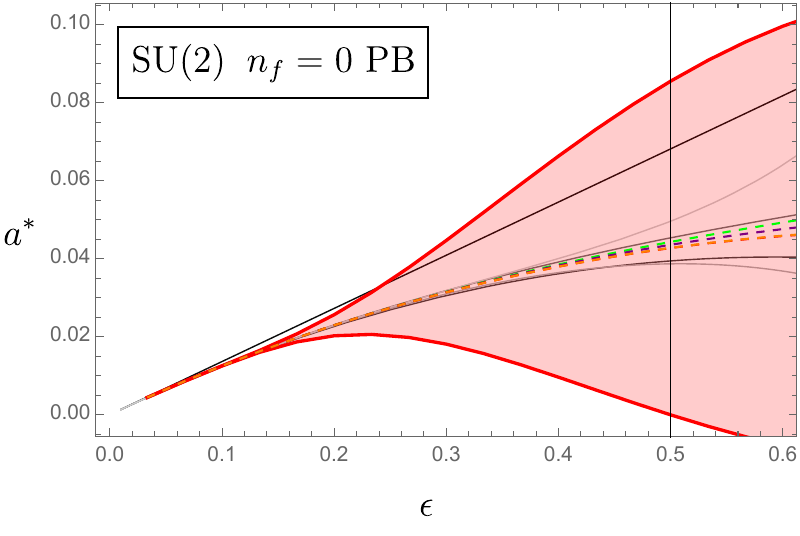} 
\raisebox{-1.58 mm}{ \includegraphics[height=0.314\textwidth]{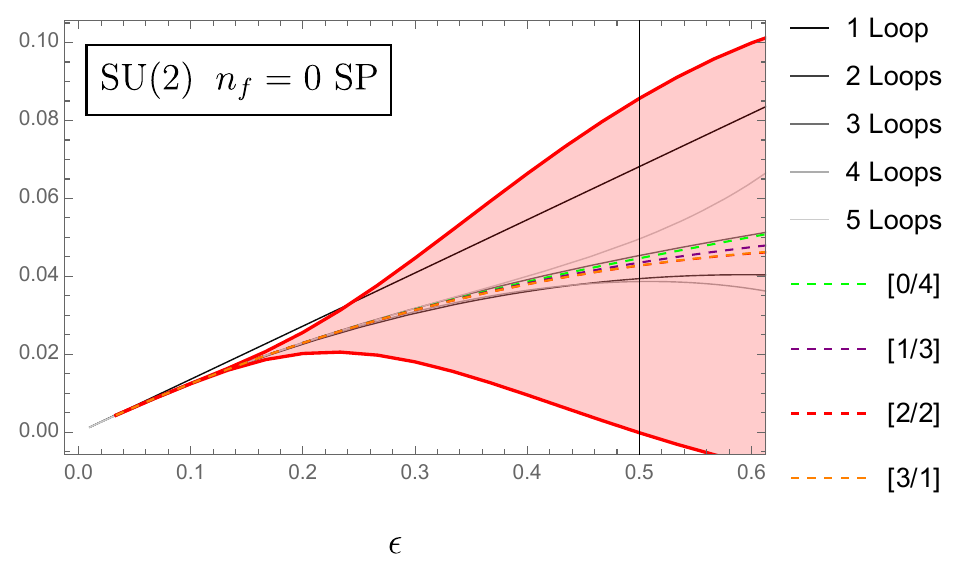}}
\caption{The value of the coupling at the fixed point $a^*$ as a function of $\epsilon$  for pure $SU(2)$ gauge theory in $d=4+2\epsilon$. (Left) Results of perturbation theory (grey lines) and of Pad\'e-Borel (PB) approximation (dashed lines). (Right) Results of perturbation theory (grey lines) and of Simple Pad\'e (SP) approximants (dashed lines). The shaded area in both panels corresponds to the  error band associated to the optimal approximant [2/2], whose central value is given by the red dashed line.}
 \label{fig:pure_su2}
\end{figure}	

\section{Results}
\label{sec:results}

We report in this section our results starting from the case of pure $SU(2)$ gauge theory, our best example, and then we generalise to different values of $n_c$ and $n_f$.
We will consider fermion matter in the fundamental representation, but clearly other choices could be investigated too. 

\subsection{Existence of a fixed point in $d=5$ for pure $SU(2)$}
\label{subsec:su2}

We report in the left panel of fig.\ref{fig:pure_su2} the value of the fixed point coupling $a^*$ as a function of $\epsilon$ obtained with both simple perturbation theory and Pad\'e-Borel approximation. In order not to clutter the picture, the error band is shown only for the optimal approximant [2/2], which has been determined 
by using exact ${\cal O}(1/n_f)$ results for $\beta$ in the  large-$n_f$ limit. See appendix \ref{appendix} for the details of this procedure. 
We see that up to five-loops every order in perturbation theory would predict a fixed point at $\epsilon = 1/2$, i.e. $d=5$ dimensions. On the other hand the values of $a^*$ differ substantially from order to order, an indication that the $\epsilon$-expansion is not convergent there. Note how Pad\'e-Borel techniques give more consistent results than the loop expansions. As the error band goes barely below zero, we expect the fixed point to exist, even if we cannot draw a strong conclusion. For illustration of the nice agreement between Pad\'e-Borel and Pad\'e methods  made at the end of section \ref{sec:stage}, we report in the right panel of fig.\ref{fig:pure_su2} the results obtained using simple Pad\'e approximation, together again with those coming from perturbation theory. The good agreement among the two results is evident, being the two panels almost indistinguishable.

\begin{figure} [t!]
\centering 
\includegraphics[width=0.7\columnwidth]{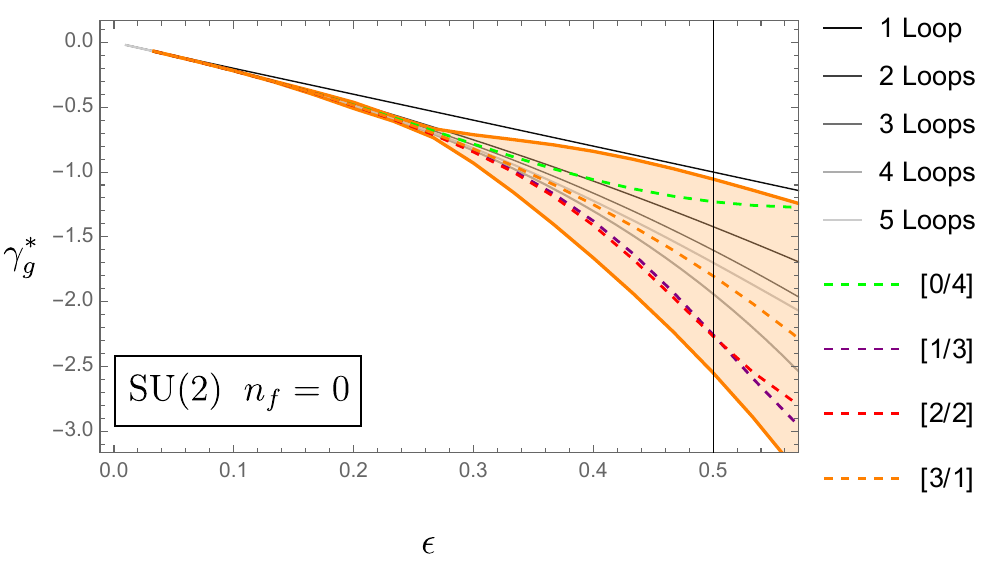}  
\caption{The value of the anomalous dimension $\gamma_g^*$ in $d=4+2\epsilon$ at the fixed point for pure $SU(2)$ gauge theories. Grey lines correspond to perturbative results, while dashed lines refer to Pad\'e-Borel approximations. The orange shaded area represents the error band for the [3/1] approximant, whose central value is the orange dashed line.} 
\label{fig:gammag_su2}
\end{figure}

In fig.\ref{fig:gammag_su2} we report the anomalous dimension $\gamma_g^*$ of the gauge kinetic operator ${\rm tr} [F_{\mu\nu}F^{\mu\nu}]$ defined as 
\be
\Delta_{F^2}=d+\gamma_g^*\,,
\ee
as a function of $\epsilon$.
As for $a^*$, the error band is shown only for the optimal approximant [3/1], determined again by using exact ${\cal O}(1/n_f)$ results for $\gamma_g$ in the  large-$n_f$ limit and excluding Pad\'e-Borel approximants with poles on the positive real axis.
For $\epsilon = 1/2$, $\gamma_g^*$ is quite large and negative, but above the unitarity bound for scalar operators $\Delta\ge(d-2)/2$, which corresponds to
\be
\gamma_g^*> -3-\epsilon \,
\ee
in $d=4+2\epsilon$ dimensions. The value of $a^*(\epsilon=1/2)\sim 4\times 10^{-2}$ could naively led us to believe that the putative UV fixed point is relatively weakly coupled.
This number is however renormalization scheme dependent and per se not that relevant, in contrast to $\gamma_g^*$ which is a scheme-independent observable.
The large value of $\gamma_g^*\sim -2$ points instead towards a putative strongly coupled fixed point. As a comparison, we note that a IR fixed point with a similar value of $a^*$
was found in ordinary $4d$ QCD with $n_f=12$ fundamental fermion flavours using the same resummation techniques (and the same scheme $\overline{\rm MS}$) used here, but resulted in values of $\gamma_g^*$ 
roughly one order of magnitude smaller \cite{DiPietro:2020jne}.
While the main source of error in the study of the QCD conformal window was given from the numerical reconstruction of the Borel function, in the $5d$ case analyzed here this is the case only for $\gamma_g$, as the dominant error for $a^*$
comes from the non-perturbative term, eq.\eqref{eqn:51}.\footnote{{Note that removing the non-perturbative contribution to the error of $a^*$ would make the error bands much smaller, providing stronger evidence for the existence of the fixed point. }}

As can be seen from fig.\ref{fig:gammag_su2}, the operator ${\rm tr}[ F_{\mu\nu}F^{\mu\nu}]$ is strongly relevant in the UV and is in fact the relevant deformation driving the UV CFT to the $SU(2)$ pure Yang-Mills theory. Whether this is the only possible relevant singlet deformation of the UV theory remains an open question.

\begin{figure} [t!]
 \centering
 \includegraphics[width=.47\columnwidth]{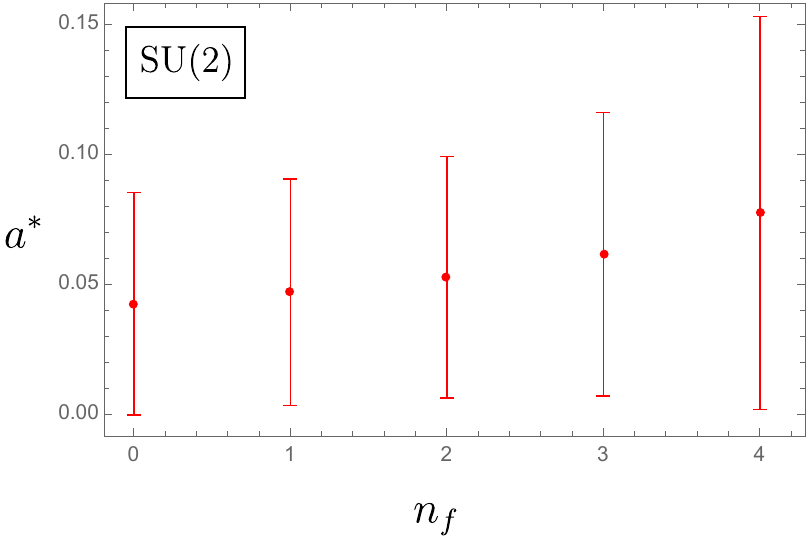}  
 \hspace*{4pt}
  \includegraphics[width=.47\columnwidth]{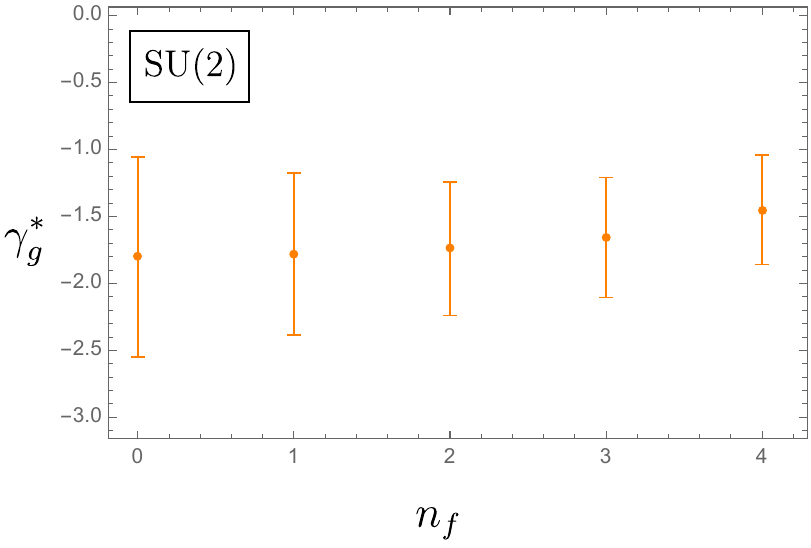}  
\caption{The value of the coupling at the fixed point $a^*$ (left) and of the anomalous dimension $\gamma_g^*$ (right) as a function of $n_f$ for $SU(2)$ gauge theories  at $\epsilon=1/2$ ($d=5$).
The central values and the corresponding error bands are obtained using the best Pad\'e-Borel approximants, respectively  [2/2] and [3/1].  }\label{fig:su2nf24}
\end{figure}

\subsection{Generalization to different values of $n_f$ and $n_c$}

The results reported in subsection \ref{subsec:su2} are easily generalized in presence of fermion matter and for other simple gauge groups. Again, the Pad\'e-Borel technique provides particularly similar results to those obtained with simple Pad\'e approximation, which we will not report in this case.
 
Let us first consider the addition of $n_f$ Dirac fermion fields in the fundamental representation coupled to $SU(2)$ gauge bosons.
In the left panel of fig.\ref{fig:su2nf24} we report the value of the fixed point coupling $a^*$ as a function of $n_f$. 
By increasing the number of fermion fields the value of $a^*$ increases. This is simple to understand by noting that the one-loop fixed point is given by
\be
a^*_{{\rm 1-loop}} = \frac{\epsilon}{\beta_0} \,.
\label{eq:a*1loop}
\ee
Since, at fixed $n_c$,  $\beta_0$ decreases as $n_f$ increases, we expect $a^*$ to increase, at least for sufficiently small $\epsilon$.  
 \begin{table} [t!]
 \centering
  \begin{tabular}{lccccc}
    \hline
    $n_f$ & 0 & 1 & 2 & 3 & 4\\
      \hline    
    $\gamma_g^*$ &-1.80(75) &-1.78(61)  &  -1.74(50) & -1.66(45)  &  -1.45(41)\\
    \hline    
    $\gamma^*$ & ---  &-0.47(12)  &  -0.52(17) & -0.59(27)  &  -0.69(46)\\
    \hline 
   \end{tabular}
   \caption{Values of the anomalous dimensions $\gamma_g^*$ and $\gamma^*$ for $SU(2)$ gauge theory with $n_f$ Dirac fermion fields in the fundamental representation at $\epsilon=1/2$ ($d=5$). The central values
are obtained averaging over the Pad\'e-Borel approximants without poles in the real positive axis and well behaved in the large-$n_f$ limit. The error band is obtained combining in quadrature the errors related to each approximant.} 
   \label{Tab:SU2}
\end{table}

The dominant source of error comes from the non-perturbative contribution, which is proportional to {$\exp(-1/(\beta_0 a^*))$}. This is 
independent of $n_f$ and proportional to {$\exp(-1/\epsilon)$}  if we use the leading 1-loop order result \eqref{eq:a*1loop} for $a^*$.
The non-perturbative source of error is actually smaller, because the resummed value of $a^*$ is typically smaller than $a^*_{{\rm 1-loop}}$.
However, the value of $a^*$ gets closer and closer to $a^*_{{\rm 1-loop}}$ as $n_f$ increases and becomes even larger at some point. This explains why the error bands
increases with $n_f$. With our choice of the error the largest value of $n_f$ for which we can confirm the existence of a UV fixed point is $n_f=4$. This result depends on the choice of the non-perturbative error; as we believe to have been quite conservative, we conclude that the maximum value $n_f^*$ for which the theory admits a UV fixed point is expected to be in the range
\begin{equation}
4\leq n_f^* \leq 10~,
\end{equation}
where the upper bound comes from the requirement of asymptotic freedom in $d=4$.

\begin{figure} [t!]
 \centering
 \includegraphics[width=.49\columnwidth]{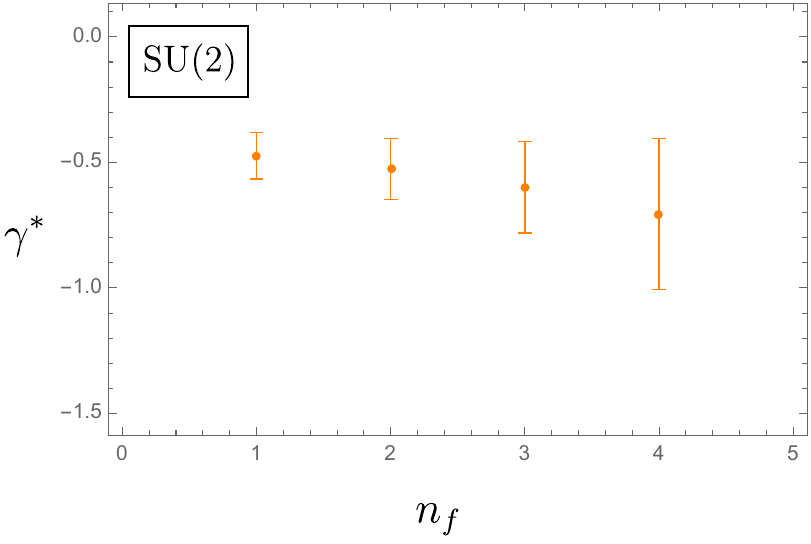}  
\caption{Value of the anomalous dimension $\gamma^*$ for $SU(2)$ as a function of $n_f$ at $\epsilon=1/2$ ($d=5$). The central values and the error bands are obtained using the best Pad\'e-Borel approximant [3/1]. } 
\label{fig:gamma_su2_nf24}
\end{figure}

In the right panel of fig.\ref{fig:su2nf24} we show $\gamma_g^*$ as a function of $n_f$.
In fig.\ref{fig:gamma_su2_nf24} we report the anomalous dimension $\gamma^*$  for the fermion mass operator defined as 
\be
\Delta(\bar{\psi}\psi)=d-1+\gamma^*\,,
\ee
as a function of $n_f$, obtained using the optimal Pad\'e-Borel approximant. By increasing the number of fermions, a larger value of $|\gamma^*|$ is found, as expected since the fixed point is more and more strongly coupled.
The unitarity bound requires in this case
\be
\gamma^*> -2-\epsilon \,,
\ee
and is always well satisfied, taking also into account the error. Perturbation theory is more stable in this case, not deviating much  from the Pad\'e-Borel approximations, at least in the range of interest.  
Again, the values are compatible with the unitarity bound. 

In tab.\ref{Tab:SU2} we report the results for $\gamma^*$ and $\gamma_g^*$ at $n_f=1,2,3,4$  obtained averaging  all well-behaved Pad\'e-Borel approximants weighted with the individual errors $\sigma_i$:  \begin{equation}
\gamma^*= \sqrt{\frac{\sum_i(\gamma_{i}^*\sigma_i^{-2})}{\sum_{i}\sigma_i^{-2}}}, \quad \quad
 \sigma=\sqrt{{\sum_{i}\sigma_i^{2}}}\ ,
 \label{eqn:gammagAverage}	
 \end{equation}
 and the same for $\gamma^*_g$.

\begin{figure} [t!]
 \centering
  \includegraphics[width=.467\columnwidth]{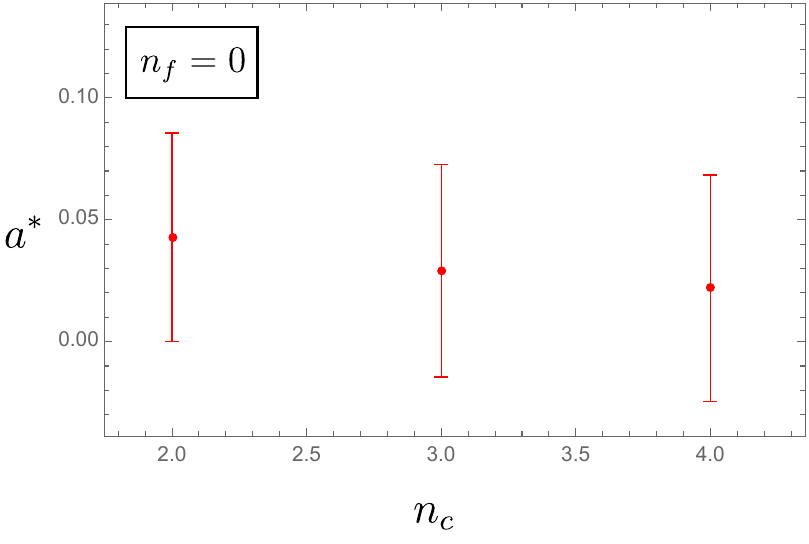} 
 \hspace*{4pt}
 \includegraphics[width=.467\columnwidth]{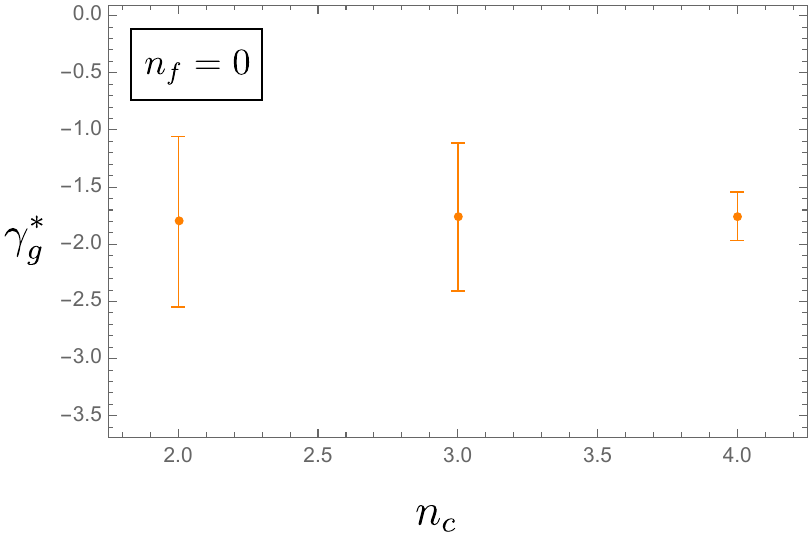}  
\caption{The value of the coupling at the fixed point $a^*$ (left) and of the anomalous dimension $\gamma_g^*$ (right) as a function of $n_c$ for pure ${SU}(n_c)$ gauge theories  at $\epsilon=1/2$ ($d=5$). The central values and the error bands are obtained using the best Pad\'e-Borel approximants, respectively  [2/2] and [3/1].  } 
\label{fig:sun_c}
\end{figure}

Let us now consider pure Yang-Mills theories with $n_c>2$. We report in the left panel of fig.\ref{fig:sun_c} the value of the fixed point $a^*$ as a function of $n_c$. The value of the fixed point decreases as the number of colours increases, while the error band, which is mostly composed by the non-perturbative contribution, keeps almost the same size. This is again simple to understand using \eqref{eq:a*1loop}.
Given that $\beta_0$ increases linearly with the rank of the gauge group, we expect $a^*$ to decrease, at least for sufficiently small $\epsilon$. 

The dominant source of error comes again from the non-perturbative contribution. In contrast to the case where we vary $n_f$ at fixed $n_c$, 
this error is approximately constant as $n_c$ varies, since so is $|a^*-a^*_{{\rm 1-loop}}|$. For $n_c\ge3$, even though the central value remains positive, the error bars in our estimate cross distinctly zero, so we cannot draw a definite conclusion.
As in the  $SU(2)$ case, adding fermion matter makes the fixed point more strongly coupled.
In the right panel of fig.\ref{fig:sun_c} we report  $\gamma_g^*$ as a function of $n_c$ using the optimal Pad\'e-Borel approximant and in tab.\ref{Tab:SUn} 
using the weighted average \eqref{eqn:gammagAverage} among all the well-behaved Pad\'e-Borel approximants.
The stability of $\gamma_g^*$ as $n_c$ varies is to some extent expected, since the first three terms of the perturbative series are independent of $n_c$. It is 
 nevertheless curious that the central values are almost %\FDC{c'era un errore nel valore central per $n_c=2$}
 identical for $n_c=2,3,4$.
This is most likely a numerical coincidence, but more speculatively one could conjecture that these CFTs are all indistinguishable at the level of local operators.\footnote{These theories are anyhow expected to be different at the level of line operators, since they have different discrete one-form global symmetries, independently of the global structure of the gauge group.}
 
The existence of a UV fixed point can also be analyzed in the Veneziano limit,  $n_f,n_c\rightarrow \infty$, $a\rightarrow 0$,  with $x=n_f/n_c$ and $\lambda_V=an_c$ held fixed. 
We omit the details of this analysis and report only the main conclusion, which 
is the evidence of a fixed point in the range {$0\leq x \leq 2.6$}. Large values for $\gamma^*$ and $\gamma_g^*$ are also found in this case, suggesting the presence of strongly coupled CFTs. The unitarity bound is well satisfied in the whole interval.

\section{Other approaches: a brief overview}
\label{sec:overview} 

 We briefly review in this subsection the main studies developed up to now on the existence of a UV stable fixed point in $5d$ non-abelian gauge theories. 
 
As mentioned, our study is an extension of the previous work \cite{Morris_2005}. This reference applied optimal truncation to the series in $\epsilon$ obtained from the then-available four-loops $\beta$-function and suggested evidence for the existence of a $5d$ fixed point for pure gauge theories with $n_c\ge2$.  Optimal truncation is however reliable only when the smallest term in the series is not the last available one and it is not conclusive if the terms are too few for this to happen. This is why numerical results were reported in \cite{Morris_2005} only for $n_c=2$, in the case of $a^*$, and $n_c=2,3$, in the case of $\nu^*$, which is an index related to the gauge anomalous dimension by the expression $\nu^*=-2/\gamma^*_g$. Such results are in agreement with ours, taken into account that we were more conservative in the estimate of the error.\footnote{Note that in ref.\cite{Morris_2005} a different notation was chosen: in order to properly compare the results one should rescale our value of $a^*$ by a factor 4.}

Let us now present other methods used to address the problem. Ref. \cite{Gies_2003} applied the exact RG equations to a certain truncated set of higher dimensional operators in $SU(n_c)$ YM, and found that the critical dimension above which the fixed point disappears is $d_{\mathrm{cr}}>5$ for $n_c=2,3,5$. 
The result seems to indicate the existence of a continuum limit in five dimensions, however the conclusion is based on the truncation of the flow equations whose reliability is hard to assess. % rigorously. 

This problem was also approached using the lattice. Given a lattice Lagrangian that reduces to the continuum $5d$ Lagrangian at large distances, one looks for a second-order phase transition by varying the lattice couplings. If such a transition is not found, then this could be a consequence of the fact that the starting Lagrangian was not general enough, and no definite conclusion can be reached about the existence of the UV fixed point in the continuum. The simplest choice for the initial action is the Wilson plaquette in the fundamental representation. With this choice and $n_c=2,\ n_f=0$, both \cite{Creutz} and \cite{kwai} found a confinement-deconfinement phase-transition of first rather than second order. Ref. \cite{kwai} went further and analyzed also a modified action with the inclusion of the Wilson plaquette in the adjoint representation. Again, in the region of the coupling space that they managed to explore they only found a first order transition, though a weaker one as the fundamental coupling was increased. The problem was revisited almost 30 years later in the recent paper \cite{Florio:2021uoz}. This reference extended the analysis of \cite{kwai} to a larger region of the coupling space and larger lattice size, and still found only first order transitions for the fundamental+adjoint action. They went on to consider a different lattice action with the fundamental Wilson plaquette of doubled size and observed the disappearance of a first order transition. Their preliminary extrapolations indicate that the disappearance is robust in the infinite volume limit. This suggests the existence of a second order transition. 

 \begin{table} [t!]
 \centering
  \begin{tabular}{lccc}
    \hline
    $n_c$ & 2 & 3& 4 \\
      \hline    
    $\gamma_g^*$ &-1.80(75) &-1.76(65)  &  -1.76(21) \\
    \hline 
   \end{tabular}
   \caption{Values of the anomalous dimensions $\gamma_g^*$ for Yang-Mills theory with different numbers of colors $n_c$ at $\epsilon=1/2$ ($d=5$). The central values
are obtained averaging over the Pad\'e-Borel approximants without poles in the real positive axis and well behaved in the large-$n_f$ limit. The error band is obtained combining in quadrature the errors related to each approximant.} 
   \label{Tab:SUn}
\end{table}
 
A different non-perturbative approach is the numerical conformal bootstrap \cite{Poland:2018epd}. Finding interacting non-supersymmetric CFTs in $d\geq 4$ with the conformal bootstrap is generally hard (see e.g. \cite{Karateev:2019pvw} for a recent attempt in $d=4$). Moreover one cannot easily input that the CFT to be looked for should be related to a non-abelian gauge theory. A method to try to target conformal gauge theories in $4d$ and $5d$ has been put forward in \cite{Li:2020bnb}: the idea is to consider the bootstrap bounds for the four-point function of $SO(N)$ vectors and look for families of kinks that are conjecturally associated to the four-point function of flavor-adjoint fermionic bilinears in the conformal gauge theory with fermionic matter.  In the specific application to $5d$, \cite{Poland:2018epd} found such a family of kinks only for relatively large values of $N$. At the moment there is neither enough evidence that such kinks are associated to CFTs 
nor a definite prediction from the conformal bootstrap for the would-be scaling dimension of the leading irrelevant operator at this kink. %It would be nice to attempt such a comparison in the future. 
The approach proposed in \cite{Poland:2018epd} is based on the flavor symmetry and therefore it is not suited to the pure YM case. For the latter, a naive possibility is to consider simply the four-point function of identical scalar operators ${\cal O}$, bounding the dimension of the operators appearing in their OPE coefficients, with the hope of finding features corresponding to the UV completion of YM, in which ${\cal O}$ can be identified with ${\rm tr}[ F_{\mu\nu}F^{\mu\nu}]$. In absence of a selection rule (such as a $\mathbb{Z}_2$ symmetry) forbidding to ${\cal O}$ itself to appear in the OPE, this vanilla bootstrap bound cannot succeed without further assumptions on the spectrum, because the interesting CFT, if it exists, would sit well within the allowed region, below the generalized free theory line.  Further predictions from $\epsilon$-expansion or other methods might help in narrowing down the appropriate gap assumptions to be made.

Ref.s \cite{BenettiGenolini:2019zth, Bertolini:2021cew} provide concrete attempts to realize the UV completion of $5d$ $SU(2)$ YM as a second-order transition in continuum QFT.  The starting point is the UV completion of the $SU(2)$ supersymmetric YM theory (SYM), the interacting super-conformal field theory known as $E_1$ theory \cite{Seiberg:1996bd}. These references consider deforming $E_1$ by both the supersymmetric deformation that leads to SYM, and by a particular supersymmetry-breaking relevant deformation. For small values of the supersymmetry breaking deformation the theory is calculable and it can be shown to flow to YM in the deep IR. It was shown in \cite{BenettiGenolini:2019zth} that certain contact terms in the correlation functions of global symmetry currents depend on the sign of the supersymmetric deformation. This signals that a transition of some kind must happen in the un-calculable region in which the supersymmetric deformation is smaller than the non-supersymmetric one. However at the moment there is no conclusive argument that this phase transition is of second order. Ref. \cite{Bertolini:2021cew} further showed an instability at infinite (bare) gauge coupling when the supersymmetry-breaking deformation is turned on, implying the existence of an intermediate phase between the two YM phases with different contact terms.

Ref. \cite{Gracey:2015xmw} proposes a realization of the UV fixed point of non-abelian gauge theories in $4<d<6$ as the IR fixed point of an RG starting from a non-unitary free theory with a higher-derivative kinetic term. This RG becomes weakly coupled in $6-\epsilon$ expansion. However this alternative description is only known to be valid in the limit of a large number of flavors, in which the fixed point in $4<d<6$ only exists for negative $g^2$.

\section{Conclusions} 
\label{sec:conclu}

We studied the extrapolation of UV fixed points of non-abelian gauge theories from $d=4+2\epsilon$ to $d=5$ using Borel-Pad\'e resummation techniques. {Our main result is illustrated in fig.s \ref{fig:pure_su2} and \ref{fig:su2nf24}%and \ref{fig:sun_c}
, where we found evidence for $5d$ fixed points for the $SU(2)$ gauge theory with $n_f \leq 4$.}%, and for the pure $SU(n_c)$ theory for $n_c \leq 4$. 
 We also used the method to provide a prediction for the dimension of the leading relevant operator at those fixed points, though the estimated relative error is typically rather large, see fig.s \ref{fig:su2nf24} and \ref{fig:sun_c}. These numbers can still provide a useful benchmark for future studies with non-perturbative techniques such as the lattice or the conformal bootstrap.

In the future, it would be interesting to study additional observables of the fixed point in $4+2\epsilon$ dimensions. These can be used to estimate observables of the conjectural $5d$ fixed point through the extrapolation to $\epsilon\to1/2$. Obvious observables to consider are the scaling dimensions of heavier operators. One-loop anomalous dimensions of some higher-dimensional operators in $4d$ Yang-Mills have been computed in \cite{Gracey:2002he}. In addition to local operators, one can also consider observables associated to the line operators of Yang-Mills theory, such as the coefficient $h$ in the one-point function of the stress-tensor in the presence of the line \cite{Kapustin:2005py}, or the Bremsstrahlung function \cite{Correa:2012at}.

Another direction for the future is to try to test the proposal of ref. \cite{BenettiGenolini:2019zth, Bertolini:2021cew} using the $F$-theorem and the $\epsilon$-expansion. A definite prediction of this proposal is the existence of an RG flow connecting the $E_1$ SCFT to the non-supersymmetric CFT that provides the UV completion of $SU(2)$ YM. Therefore according to the $F$-theorem \cite{Klebanov:2011gs} (which is still a conjecture in $5d$) the five-sphere partition function coefficient $F_{E_1}$ of the SCFT and that of the non-supersymmetric CFT $F_{\text{CFT}}$ should satisfy $F_{E_1} > F_{\text{CFT}}$. The quantity $F_{E_1}$ has been computed using localization in \cite{Chang:2017cdx}. Assuming the CFT is the continuation of the $4+2\epsilon$ fixed point, $F_{\text{CFT}}$ can be computed in perturbation theory using the technique of \cite{Fei:2015oha, Giombi:2015haa}. If the extrapolation to $\epsilon \to 1/2$ gives a result of the same order and smaller than $F_{E_1}$, this would be an indication in favor of the proposal.

\section*{Acknowledgments}

We thank M. Bertolini for useful discussions. LD and MS are partially supported by INFN Iniziativa Specifica ST\&FI. LD also acknowledges support by the program ``Rita Levi Montalcini'' for young researchers.

\appendix

\section{Large $n_f$, Selection of Pad\'e-Borel Approximants and Estimate of the Error}
\label{appendix}

The choice of the Pad\'e-Borel (and ordinary Pad\'e) approximant for a given truncated sum is not univocal.
 We show in this appendix, following \cite{DiPietro:2020jne}, how one can use exact results in the large-$n_f$ limit to test the accuracy of different approximants and possibly select optimal ones.  
To this aim  we can compare the results obtained with the known exact functions $\beta^{(1)}$ and $\gamma^{(1)}$ at large $n_f$ with the approximations found with Pad\'e-Borel resummation.

\begin{figure}[t!]
 \centering
 \includegraphics[width=0.65\columnwidth]{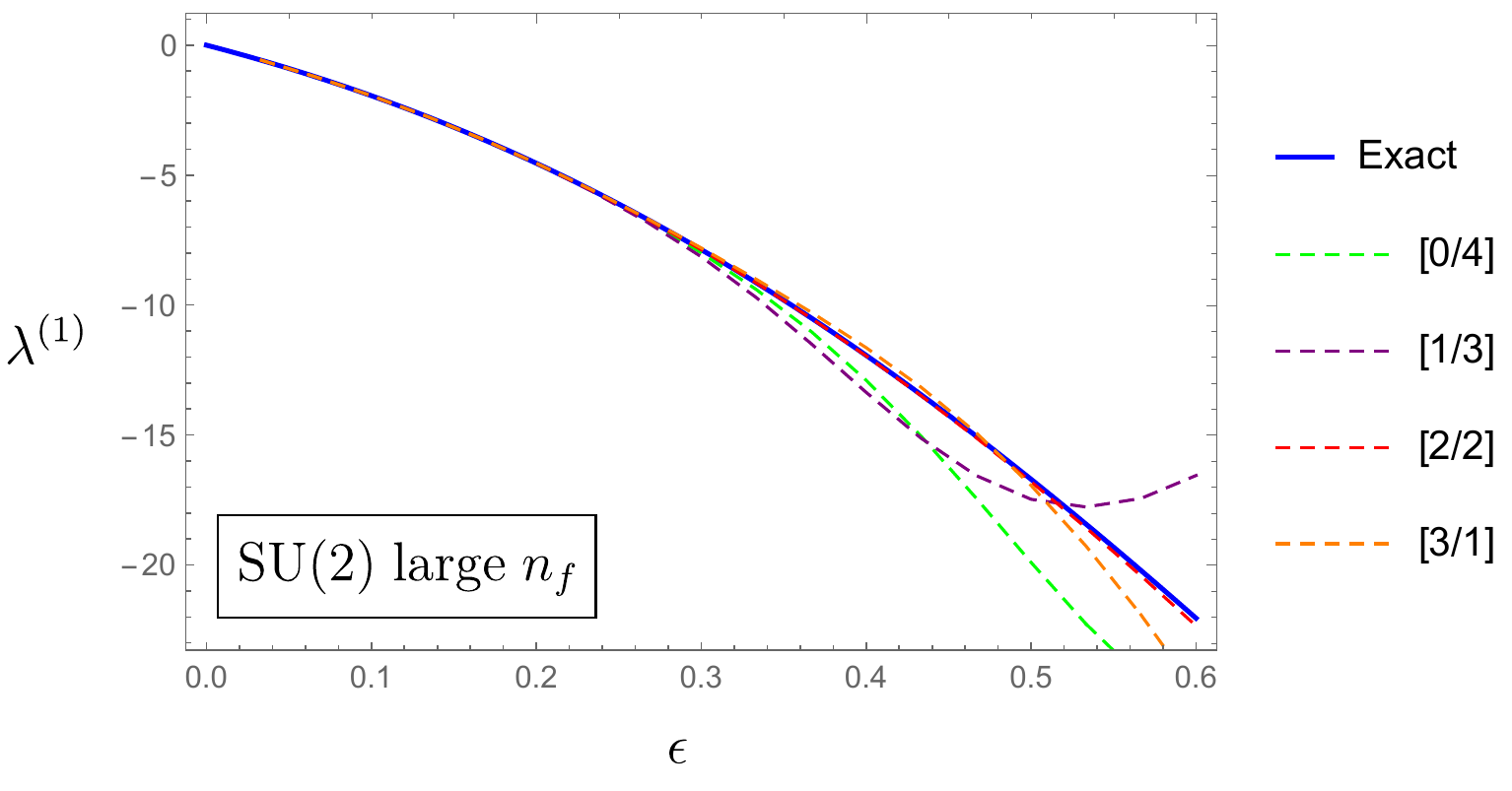} 
\caption{Comparison of $\lambda^{(1)}$ as a function of $\epsilon$  for $SU(2)$ gauge theory. 
 The continuous blue line denotes the exact result obtained using large-$n_f$ methods, the dashed lines denote the Borel resummation using Pad\'e-Borel approximants as indicated in the legend.}
 \label{fig:largenf} 
 \end{figure}

The large-$n_f$ limit is defined as $n_f\rightarrow \infty$, $a\rightarrow 0$, with $n_c$ and the 't Hooft-like coupling $\lambda = n_f a$ held fixed. In this limit \eqref{eqn:46} becomes
\begin{equation}
\beta(\lambda)=\epsilon\lambda+\frac{2}{3}\lambda^2+\frac{1}{n_f}\beta^{(1)}(\lambda)+\mathcal{O}\biggl(\frac{1}{n_f^2}\biggr)\,,
\label{eqn:50b}
\end{equation}
with $\beta^{(1)}$ a known function \cite{Gracey:1996he}, analytic at $\lambda = 0$ (see e.g. \cite{Holdom:2010qs} for a particularly nice explicit form).
Correspondingly, \eqref{eqn:48} turns into
\begin{equation}
\lambda^*(\epsilon)=-\frac{3}{2}\epsilon+\frac{1}{n_f}\lambda^{(1)}(\epsilon)+\mathcal{O}\biggl(\frac{1}{n_f^2}\biggr)\,, \quad \lambda^{(1)}(\epsilon) = \frac{1}{\epsilon}\beta^{(1)}\left(-\frac{3}{2}\epsilon\right).
\label{eqn:50}
\end{equation}
As expected, the fixed point turns negative for large $n_f$ and is unphysical. 
 In this regime we choose the Pad\'e-Borel approximant that better reproduces the exact result and assume that this
 remains valid at finite $n_f$, when the fixed point is possibly physical. It is clear from fig.\ref{fig:largenf} that the [2/2] approximant is the one that better matches the exact function, at least in the interval of interest.  We have reported the result for $SU(2)$, but the same applies for $n_c=3,4$.
 
We can use a similar procedure for $\gamma^*$ and $\gamma_g^*$. In the large-$n_f$ limit the expression for $\gamma$ can be expanded as
\begin{equation} 
\gamma^*(\epsilon)=\frac{1}{n_f}\gamma^{(1)}\left(-\frac{3}{2}\epsilon\right)+\mathcal{O}\biggl(\frac{1}{n_f^2}\biggr)\,,
\end{equation}
with $\gamma^{(1)}$, like $\beta^{(1)}$, a known function analytic at $\lambda = 0$ \cite{Espriu:1982pb,PalanquesMestre:1983zy}.
For $\gamma_g$ we have
\begin{equation} 
\gamma_g^*(\epsilon)=2\frac{\partial \beta(\lambda)}{\partial\lambda}\bigg|_{\lambda=\lambda^*(\epsilon)}=-2\epsilon+\frac{1}{n_f}\gamma_g^{(1)}\left(-\frac{3}{2}\epsilon\right)+\mathcal{O}\biggl(\frac{1}{n_f^2}\biggr)\,, 
\end{equation}
with $\gamma^{(1)}_g$ obtained deriving eq.\eqref{eqn:50b} and replacing the fixed point of eq.\eqref{eqn:50}. In the case of $\gamma^{(1)}$ the comparison to the exact function does not lead to a clear choice of an approximant with respect to the others: a good matching is obtained with [3/1], [2/2] and [1/3] Pad\'e-Borel approximants. None of them has poles in the real positive axis. The comparison for $\gamma_g^{(1)}$ leads to the same considerations found for $\lambda^{(1)}$, but in this case only the [3/1] can be selected, the [2/2] having poles at finite $n_f$.

We now review the procedure used for the estimate of the error. Let us consider the function $f$ with
a divergent and non-Borel resummable asymptotic expansion:
\begin{equation}
f(t)\sim t^{\zeta_f}\sum_{n=0}^\infty f_n t^n,
\label{eqn:10}
\end{equation}
where $\zeta_f$ is chosen by requiring that the coefficient $f_0$ is non-null.  The Pad\'e-Borel approximation is defined as
\begin{equation}
f_B^{[m/n]}(t)=t^{\zeta_f}\int_0^\infty dz\ z^be^{-z}\mathcal{B}_b^{[m/n]}(z t),
\label{eqn:a1}
\end{equation}
where $\mathcal{B}_b^{[m/n]}$ is the $[m/n]$ Pad\'e approximation to the Borel-LeRoy transform of the series with parameter $b>-1$.
We define the total error $\Delta^{[m/n]}$ as the sum of three contributions:
\begin{equation}
\Delta^{[m/n]}=\Delta^{[m/n]}_\mathrm{conv}+\Delta^{[m/n]}_b+\Delta_\mathrm{np}.
\end{equation}
As only Pad\'e-Borel approximants without poles in the real positive axis are selected, no contributions from the residues,  denoted by 
$\Delta_r^{[m/n]}$ in \cite{DiPietro:2020jne},  enter in the error. In order to be reasonably conservative we have included a new term in the error, not present in 
\cite{DiPietro:2020jne}, which estimates the convergence of the approximants and is relevant in the present analysis. Such uncertainty, denoted by $\Delta^{[m/n]}_\mathrm{conv}$, can be estimated by computing the distance among the Pad\'e-Borel approximant and the subsequent one belonging to the same family:\footnote{In order to avoid misleading results, we have checked that the approximants used for such comparisons have small residues in the real positive axis or, even better, no poles at all.}
\begin{equation}
\Delta^{[m/n]}_\mathrm{conv}=\left|f_B^{[m/n]}-f_B^{[m-1/n-1]}\right| \,.
\end{equation}
 If the $[m-1/n-1]$ approximant is not avalaible, the $[m-1/n]$ is selected instead.\\
The term $\Delta^{[m/n]}_b$ measures the error caused by the arbitrariness on the choice of the LeRoy parameter $b$. It is indeed convenient to select a whole grid of values $\mathcal{B}=[b_0-\Delta b,\ b_0+\Delta b]$ and repeat the approximation for each of them; in our computation we choose $b_0=10$, $\Delta b=10$ and the spacing of the grid equal to 2 for $a^*$,  $\gamma_g^*$ and $\gamma^*$. The error can be then defined as
\begin{equation}
\Delta^{[m/n]}_b=\frac{1}{2}\left| \max_{b\in\mathcal{B}}f_B^{[m/n]}(b)-\min_{b\in\mathcal{B}}f_B^{[m/n]}(b)\right|.
\label{eqn:a8}
\end{equation}
The last contribution to the error is not due to the resummation technique but corresponds instead to non-perturbative corrections, which we write as 
\begin{equation}
\Delta_{\mathrm{np}}=c_{\mathrm{np}}e^{-\frac{1}{{\beta_0 a^*}}},
\label{eqn:51}
\end{equation}
where $c_\mathrm{np}$ is an arbitrary coefficient that we take {equal to 1}, $a^*$ is the value of the fixed point and { $1/\beta_0$ }is the leading term in eq.\eqref{eq:betaNP}. In all the cases analyzed here, the first instanton anti-instanton singularities provide indeed subleading non-perturbative corrections.

\bibliographystyle{JHEP}
\bibliography{5d}

\end{document}